**NUCLEI, PARTICLES, FIELDS,
GRAVITATION, AND ASTROPHYSICS**

# Magnetic Monopoles and Dark Matter

**V. V. Burdyuzha**

*Astrospace Center, Lebedev Physical Institute, Russian Academy of Sciences, Profsoyuznaya ul. 84/32, Moscow, 117997 Russia*
*e-mail: burdyuzh@asc.rssi.ru*
Received October 8, 2017; in final form, May 29, 2018

**Abstract**—Schwinger's idea about the magnetic world of the early Universe, in which magnetic charges (monopoles) and magnetic atoms ($g^+g^-$) could be formed, is developed. In the present-day Universe magnetic charges with energies in the GeV range can be formed in the magnetospheres of young pulsars in superstrong magnetic fields. Spectroscopic features of magnetic atoms and possibilities for their observations are discussed. Relic magnetic atoms can contribute up to 18% to the dark matter density. The gamma-ray excess at our Galactic center could arise under two-photon annihilation of magnetic charges as a cooperative effect from neutron stars. A sharp physical difference of Schwinger's magnetic world from Dirac's present-day electric world is pointed out. Artificial magnetic monopoles are also mentioned briefly.

## 1. INTRODUCTION

In electrodynamics the problem of magnetic charges has not been completely clarified, although the assertion that there are no free magnetic charges in nature has become fixed owing to Maxwell's classical equations. The problem has not become clearer even after the detection of structures similar to Dirac magnetic charges in laboratory conditions [1–3]. They were called artificial magnetic monopoles. Therefore, the authors of [1–3] predict a revolution in physics. In contrast, in this paper we transfer some of the "revolutionary ideas" to the cosmos. Magnetic atoms and even isolated magnetic charges that are "blown out", as electron and positrons, from young neutron stars can exist in cosmic conditions and, what is more, probably not all of the high-energy relic magnetic atoms have decayed.

Magnetic charges were first mentioned by Curie [4] more than 120 years ago. They were detected by the Austrian physicist Ehrenhaft [5] and the Soviet physicist Sizov [6]. However, nobody believed these scientists, because in Maxwell's equations div**B** = 0 and Maxwell's equations are "sacred". Sizov in his time was not even certified as a scientist, because he was concerned with "rubbish". Magnetic charges of high energies $10^{15}$–$10^{16}$ GeV were probably observed in cosmic rays by Cabrera [7]. However, there were only two events in his experiment and, what is more, these were observed on Saint Valentine's Day, which caused distrust of the physical community. And, of course, the main argument for the impossibility to observe isolated magnetic poles (monopoles) comes from the course of theoretical physics by Landau and Lifshitz [8].

However, not all of the physicists "neglected" magnetic charges, especially in the context of the early Universe. Furthermore, Sakharov [9] pointed out that black miniholes could evaporate heavy monopoles. The inflationary cosmological model was developed to avoid a great over-excess (up to 16 orders of magnitude) of high-energy GUT monopoles (GUT stands for grand unified theory). Zel'dovich and Khlopov [10] showed that the present-day concentration of relic monopoles with energies in the TeV range is extremely low ($10^{-19}$ cm$^{-3}$). Schwinger published the review "A Magnetic Model of Matter" in UFN [11], thereby predicting the magnetic world of the early Universe. In addition, an interesting remark was made in [12]: "monopoles cannot play any role in the Standard Model, and in its usual extensions, up to the Planck scale, on which they can lead to space discreteness."

Here we want to draw attention to the possibility of detecting monopoles with energies in the GeV range in cosmic conditions and to enhance the role of high-energy magnetic monopoles in the early Universe. The main reason for our desire to revisit the leptogenesis is a huge magnitude of magnetic forces. In the symmetric (Schwinger) case, the magnetic forces are stronger than the electric ones approximately by a factor of ~20000 (Section 3). The question about the influence of these forces on the generation of baryon asymmetry of the Universe immediately arises here, because all of the known effects leading to CP-symmetry breaking are weak.

The detection of artificial magnetic charges in spin ice as a result of geometrical (magnetic) frustration is actually a very interesting event emitting Dirac mono-poles. The deconfinement of effective magnetic charges in a crystal lattice (to be more precise, the deconfinement of zero-dimension point topological defects) arises at temperatures close to absolute zero. Note that the spectrum of topological defects in spin systems includes vortices, solitonic vortices, skyrmions, monopoles, and knots [13]. The new term "magnetricity" (by analogy with electricity) and even such a term as "magnetolyte" (by analogy with electrolyte) were introduced for the emerged current of magnetic charges. These experiments and magnetic frustration physics are described in detail in [13–17].

In other words, for the appearance of a current of such magnetic charges the topological order in crystals is violated due to magnetic frustration [17]. As Bramwell [16], one of the ideologists of spin ice, monopoles, and magnetricity, said, "magnetricity is a current of thermally excited defects in spin ice". Possibly, it should be added to this definition that it is necessary to take into account the spin correlations. The study of magnetic systems in low-temperature physics includes several physical concepts: spin ice, magnetic monopoles, anomalous Kondo and Hall effects [18]. In the opinion of Zvyagin [18], we already observe a new physics in frustrated magnets and it is probably hard not to agree with this.

## 2. MAGNETIC CHARGES, THEIR ENERGIES, AND THEIR SEARCH

Magnetic monopoles have been and are being searched for in various energy ranges from $10^{16}$ GeV to a few GeV or even lower and, of course, their search is conducted by various methods. At the Large Hadron Collider this is the MoEDAL experiment. A brief theory of leptonic magnetic monopoles is presented in [19, 20] and it was shown that a light magnetic monopole could be included in a consistent way in the Standard Model through the extension of the leptonic sector, i.e., a magnetic analog of the Standard Model has been created. Leptonic magnetic monopoles can be focused. A special accelerator is being built in France for this purpose [19]. Note that the observation of a moving magnetic monopole with charge $g = 137e$ and a mass larger than 100 proton masses was announced in [21]. It should also be noted that many papers, which make no sense to cite here, were devoted to closing the subject of the existence of magnetic charges (monopoles).

Our interest in magnetic charges is associated with the realization of a GeV monopolium ($g^+g^-$) atomic system in cosmic conditions, by analogy with positronium ($e^+e^-$) [22], in which some transitions can be observed in the gamma-ray range before annihilation, as, incidentally, in positronium, but in the case of positronium this is the millimeter and radio bands. Furthermore, in principle, it is possible to detect isolated magnetic charges in cosmic experiments onboard the International Space Station, but this question requires an additional study. Magnetic charges (monopoles) and atoms consisting of them should be included in the composition of dark matter. Our explanation of the gamma-ray excess at the Galactic center in the energy range 1–3 GeV is the combined effect from the annihilation of produced magnetic charges in the magnetospheres of a large number of young neutron stars—pulsars (this hypothesis will be discussed in Section 5).

Here we will focus our attention on magnetic monopoles with energies in the GeV range. Monopoles of very high energies ($10^{15}$–$10^{16}$ GeV), of course, will be investigated also, but more briefly. Their detection is envisaged in the Dubna experiment on Lake Baikal [23, 24] and in many other experiments worldwide. Sullivan and Fryberger [25] consider the execution of an experiment in Japan by the BELLE II Collaboration at the KEK facility aimed at searching for magnetic monopoles with a mass of 4–5 GeV/$c^2$ for the natural, in their opinion, case where the electric and magnetic charges are equal to each other, i.e., $e = g$.

## 3. PHYSICAL SUBSTANTIATION OF THE PRESENCE OF MAGNETIC CHARGES

Formally, Maxwell's classical equations do not suggest a complete symmetry of electric and magnetic processes and these equations yield correct results, although the presence of magnetic charges ($g$) can explain the electric charge quantization. An important dependence was derived in his time by Dirac [26]:

$$\frac{eg}{\hbar c} = \frac{k}{2}, \quad k = 0, \pm 1, \pm 2, \pm 3, \ldots \quad (1)$$

($k$ is the monopole quantum number). This classical definition of $k$ differs from its quantum definition given in the review [11]. In his time Dirac accepted the challenge of Curie [4] and suggested the existence of an elementary magnetic charge. The relation between the charges $g = 68.5e$ follows from the condition (1) at $k = 1$, i.e., the magnetic charge is very large and this is its main peculiarity. Furthermore, we know well that the fine-structure constant $\alpha_e = e^2/\hbar c = 1/137$ characterizes the force of attraction (or repulsion) between two electric charges. Accordingly, $\alpha_g = g^2/\hbar c = 34.25$ will characterize the force of attraction (or repulsion) between two magnetic charges. The ratio of these two constants is

$$\frac{g^2/\hbar c}{e^2/\hbar c} = 4692.25.$$

Since $\alpha_g \gg 1$, accurate quantum-mechanical calculations of the level structure of magnetic atoms ($g^+g^-$) cannot be made. Schwinger [11] hypothesized that the coefficient $k$ in Eq. (1) could take only even values.

For $k = 2$ we will then have $g = 137e$ and the ratio of the fine-structure constants

$$\frac{g^2/\hbar c}{e^2/\hbar c} = 18769$$

is larger than that in Dirac's case by a factor of 4. The spectroscopy of the more symmetric magnetic world will differ radically from the spectroscopy of our (Dirac) world. The huge ratio of these constants could not but affect the physical processes in the early Universe, when magnetic monopoles were formed. Of course, magnetic charges immediately after their production were bound due to a very strong magnetic interaction into the simplest atomic system, monopolium ($g^+g^-$), whose spectral features should be discussed and we should consider how they can be observed.

In [26] Dirac wrote out his famous relativistic equation in the form

$$H^2 \psi = (p^2 + m^2) \psi, \quad (2)$$

whose structure suggests the presence of a second particle. As it turned out later, this led to the detection of the positron. However, these could also be magnetic charges of different signs. Here it is pertinent to note two more points. First, quite long ago Schwinger [11] drew attention to the possible existence of a new dual particle (dion) that has both electric, $(-1/3)e$, and magnetic, $(2/3)g$, charges. The quantization condition for these two charges will then be

$$\frac{e_1 g_2 - e_2 g_1}{\hbar c} = k. \quad (3)$$

Here $k$ is an integer, while the electric ($e = 1/3$) and magnetic ($g = 2/3$) charges are fractional. Dions are particles with spin $s = 1/2$. The second remark is associated with Parker's limit [27]. The essence of this remark is that the galactic magnetic field should not change and magnetic monopoles should not reduce it when moving along field lines. Parker's limit gives a constraint on the flux of supermassive magnetic monopoles in experiments on Earth:

$$F_g < 3 \times 10^{-15} \text{ cm}^{-2} \text{ s}^{-1}. \quad (4)$$

The reference limit for the flux of isolated supermassive monopoles gives

$$F_g < 10^{-15} \text{ cm}^{-2} \text{ cp}^{-2} \text{ s}^{-1}. \quad (5)$$

Parker's remark could be a "blank shot", because there must be "few" isolated monopoles at the present epoch in the Universe. On the other hand, young neutron stars (gamma-ray pulsars) could "leave the footprints" of monopoles in regions with their highest concentration (for example, the Galactic center).

We reiterate once again that all cosmological heavy monopoles ($10^{15}$–$10^{16}$ GeV) in the leptogenesis period were immediately bound into a magnetic atom, monopolium, at huge redshifts $z \sim 10^{10}$–$10^{11}$, to form Schwinger's magnetic world. A monopole pair with an energy in the GeV range (along with an electron–positron one) could be formed in the magnetospheres of young pulsars in the superstrong magnetic fields of neutron stars and be blown out from them. Leptonic-mass magnetic monopoles [19, 20] are an interesting phenomenon, especially since, as has already been noted, a special accelerator is being built in France for their search. Dual particles, dions, "stay aloof" in current physics, but they have not yet been detected, although they, along with magnetic monopoles, could be "implicated" in $CP$ violation in the early Universe. We will also mention the possible existence of such particles as magnetic preons, i.e., the next level of matter already in the magnetic world.

## 4. MONOPOLES WITH GeV/$c^2$ MASSES IN THE VIEW OF DIRAC AND SCHWINGER

Dirac's theory does not predict the magnetic monopole mass, but it is often assumed that the monopole mass can be

$$\begin{aligned} m_g &= (g/e)^2 m_e \\ &= 4692.25 m_e \approx 2.56 m_p \approx 2.4 \text{ GeV}/c^2. \end{aligned} \quad (6)$$

In this case, the classical monopole radius is equal to the classical electron radius, which is probably natural:

$$r_g = g^2/m_g c^2, \quad r_e = e^2/m_e c^2. \quad (7)$$

If, however, the monopole radius is set equal to the classical proton radius ($\sim 0.8 \times 10^{-13}$ cm), then a considerably larger value of $m_g \approx 8.7 m_p$ is obtained for the monopole mass. A detailed discussion about the masses of magnetic monopoles can be found in [28]. The masses of these monopoles can lie in the range of TeV/$c^2$ or higher. The lower limit for the mass of a Dirac monopole was estimated quite long ago in [29] from the results of a $(g - 2)$ experiment: $m_g = 11 m_\mu \approx 1.2 m_p$. This $(g - 2)$ experiment was conceived for an accurate measurement of the muon magnetic moment (a new physics was searched for already in the 1960s).

Recall that monopolium, along positronium, has two systems of levels consisting of ortho- and para-modifications related to the orientation of their spins. Thus, before annihilation magnetic charges with a mass of 2.4 GeV/$c^2$ could form an atomic system—monopolium ($g^+g^-$). In this atomic system the energy of the $L_\alpha$ transition is about 1.8 GeV, while the energy of the ortho–para transition is about 282 keV. Here the energies were calculated by the similarity method with similar transitions in positronium ($e^+e^-$). It is also interesting to estimate the Bohr radius of this magnetic atom and to compare it with the Bohr radius of the hydrogen atom:

$$r_g^B = \frac{\hbar^2}{m_g g^2} \approx 10^{-12} \text{ cm},$$
$$r_H^B = \frac{\hbar^2}{m_e e^2} \approx 5 \times 10^{-9} \text{ cm}. \quad (8)$$

There are several research works on monopolium and all of them are associated with Vento's papers [20, 31], but there is also one "old" paper [32]. Note that the monopolium two-photon annihilation energy in our case is 2.4 GeV per each photon and precisely this process should be discussed. Three-photon annihilation in positronium is much less probable [16] and the situation must be similar in the case of monopolium. The production of monopoles with masses in the GeV range on accelerators is impossible, because the J/ψ particle production cross section is larger than the monopolium production cross section at least by an order of magnitude (the J/ψ particle mass is 3.1 GeV/$c^2$). This will be discussed in Section 7.

Schwinger [33] proposed to modify Dirac's quantization condition (2) in such a way that the monopole quantum number $k$ could take only even values. In this case, the minimum value is $k = 2$ and here there is some symmetry that "forces" us to go into earlier epochs of evolution of the Universe. The relation between the magnetic and electric charges then becomes $g = 137e$. In view of their exceptional importance for cosmology, we will repeat these trivial relations here:

$$\alpha_m = \frac{g^2}{\hbar c} = 137, \quad \alpha_m \alpha_e = 1,$$
$$\frac{\alpha_m}{\alpha_e} = 18769, \quad (9)$$

i.e., in Schwinger's world the situation differs radically from our electric world. The minimum mass of the magnetic charge in Schwinger's symmetric world must probably be

$$m_g = (g/e)^2 m_e$$
$$= 18769 m_e \approx 10.24 m_p \approx 9.6 \text{ GeV}/c^2. \quad (10)$$

Here, as in Dirac's case, the classical magnetic monopole radius is equal to the classical electron radius. Before the annihilation of Schwinger magnetic charges with a minimum mass of 9.6 GeV/$c^2$, they could also form an atomic system—monopolium ($g^+g^-$). In this atomic system the energy of the $L_\alpha$ transition is about 7.2 GeV, while the energy of the ortho–para transition is about 1.13 MeV. These transitions were also calculated by the similarity method with similar transitions in positronium.

Actually, there is a deeper connection between Dirac's and Schwinger's views. According to Dirac, there is no dual symmetry of Maxwell's equations. According to Schwinger, it is present. Furthermore, note once again that dions deserve greater attention, as does Schwinger's entire magnetic world of the early Universe, in which Maxwell's equations could be dually symmetric:

$$\text{curl}\mathbf{H} = \frac{1}{c}\left(\frac{\partial \mathbf{D}}{\partial t} + 4\pi j_e\right), \quad \text{div}\mathbf{D} = 4\pi\rho_e,$$
$$\text{curl}\mathbf{E} = \frac{1}{c}\left(-\frac{\partial \mathbf{B}}{\partial t} + 4\pi j_g\right), \quad \text{div}\mathbf{B} = 4\pi\rho_g. \quad (11)$$

## 5. THE GAMMA-RAY EXCESS AT THE GALACTIC CENTER—THE ANNIHILATION OF MAGNETIC CHARGES WITH ENERGIES IN THE GeV RANGE

Recently, the Fermi Gamma-Ray Space Telescope detected a gamma-ray excess in the energy range 1–3 GeV from the region surrounding the center of our Galaxy [34], which can be interesting for us if this is assumed to be the annihilation of monopolium. The observed spectrum of the gamma-ray excess is broadened up to 10 GeV and extends at an angle of 5° toward the Galactic center [35]. Furthermore, the distribution of photons in the spectrum is not smooth [36, 37]. Such a gamma-ray spectrum could probably be formed by unresolved point sources, young neutron stars (millisecond pulsars), which together give the observed gamma-ray excess, i.e., a new population of sources the gamma-ray flux from which was below the detection threshold of the Fermi detectors is predicted here. One of the preliminary attempts to explain the gamma-ray excess was to assume the annihilation of dark matter particles into Standard Model particles [38]. As it seems to us, a different physical model—the cooperative effect from the annihilation of monopolium in the magnetospheres of young gamma-ray pulsars, which is related to the inverse Compton effect, is more suitable for interpreting the gamma-ray excess at the Galactic center. Relativistic electrons exist in the magnetospheres of pulsars near the magnetic poles, moving at small pitch angles to the radial magnetic field. In this case, Compton scattering can broaden and shift the 2.4-GeV annihilation line. Assuming that $\varepsilon/E \ll 1$, where $\varepsilon$ is the energy of the incident gamma-ray photon and $E$ is the electron energy, the recoil upon scattering may be neglected [39]. In our case, $\varepsilon \approx 2.4$ GeV and the Thomson approximation can be used for electrons with $E \gg 2.4$ GeV. If the emission is concentrated in a narrow cone of pitch angles to the radial magnetic field, $\theta \gg m_e c^2/E$, then the energy of the scattered gamma-ray photon is

$$\varepsilon_{sc} \approx (1/3)\varepsilon\theta^2(E/m_e c^2)^2. \quad (12)$$

If we take $\varepsilon_{sc} \sim 4\varepsilon \sim 10$ GeV and an electron energy $E \sim 20000 m_e c^2$, then $\theta \sim 0.01°$, i.e., we assume that the scattering effects will lead to an effective broadening and shift of the $g^+g^-$ annihilation line up to 10 GeV, as is observed. The above estimates were made here for an individual gamma-ray pulsar.

It should also be noted that the density of monopoles must not close the Universe, $n_g m_g < \rho_{cr}$. In this case, the density of monopoles must be no more than $10^{-6}$ cm$^{-3}$. This remark is also true for the monopolium concentration. An observable gamma-ray flux ($10^{-7}$ photons cm$^{-2}$ s$^{-1}$) cannot be obtained without invoking an anisotropy of at least one order of magnitude. This is a weak deviation from isotropy, but it is unavoidable. This, in some sense, is a radiation pattern. Let us also dwell on the micro- and macroscopic constraints.

If the mean free path of monopoles in a plasma $\lambda > r_0$, where $r_0 = g^2/k_B T$ is the size at which the Coulomb attraction is significant, then free monopoles can be annihilated. In the case of $\lambda < r_0$, the diffusion approximation should be applied. Our estimate is $r_0 = (68.5e)^2/k_B T \approx 2 \times 10^{-12}$ cm at $E = 100$ GeV. The diffusion approximation is probably valid only in a very early Universe and it does not work at present. We assert that the gamma-ray excess is the production and annihilation of magnetic charges with energies in the GeV range in the superstrong magnetic fields of young neutron stars at $B \geq 10^{12}$ G in the Galactic center region. In this case, the luminosity under monopolium two-photon annihilation is

$$L = m_g c^2 \sigma v n_g N_g n_{ns} = (4 \times 10^{-24})(9 \times 10^{20}) \\ \times (4 \times 10^{-32}) \times 10^5 (2 \times 10^{19})(3 \times 10^{38}) \times 10^7 \quad (13) \\ = 2 \times 10^{36} \text{ erg s}^{-1},$$

where $N_g$ is the total number of pairs of magnetic charges in the magnetic column of a neutron star ($3 \times 10^{38}$), $n_{ns}$ is the number of young neutron stars ($10^7$), $v$ is the monopolium velocity, and the monopolium density is $n_g = 2 \times 10^{19}$ cm$^{-3}$. Here we used the annihilation cross section $\sigma = 4 \times 10^{-32}$ cm$^2$ calculated by us in [40]. Nevertheless, the questions related to the annihilation time of magnetic charges in the column of a young neutron star remain, but the "required" luminosity, probably, can be estimated. Such an estimation of the $e^+e^-$ annihilation line luminosity for the same column parameters and the same number of neutron stars gives

$$L = m_e c^2 \sigma v n_e N_e n_{ns} \approx 3 \times 10^{-38} \text{ erg s}^{-1}. \quad (14)$$

Note that the 511-keV positronium annihilation line was observed from the Galactic center quite long ago by Russian scientists [41].

## 6. HEAVY MAGNETIC MONPOLES WITH MASSES $10^{15}$–$10^{16}$ GeV/$c^2$

Note that the authors of [42] understood that a pair of heavy magnetic charges of opposite signs could produce ultrahigh-energy cosmic rays upon annihilation. Let us discuss the production of heavy magnetic monopoles in the early Universe. Many authors have already taken this path [10, 32, 43]. As has been noted above, the early Universe is described by the diffusion approximation, because $\lambda < r_0$. The diffusion equation for a magnetic monopole was written out in this case in [10]. It is

$$\frac{\partial n(r,t)}{\partial t} = \frac{D}{r^2} \frac{\partial}{\partial r} \left[ r^2 \left( \frac{\partial n(r,t)}{\partial r} + \frac{g^2}{r^2 k_B T} \right) n(r,t) \right], \quad (15)$$

where $D \approx (2/3)\lambda v$ is the diffusion coefficient. While investigating this equation, the authors of [10] point out that the annihilation of GUT monopoles virtually ends already at $t \sim 10^{-5}$ s and, in this case, the present-day residual density of isolated magnetic monopoles will be exceptionally low, $n_g \sim 10^{-19}$ cm$^{-3}$. This conclusion reached by the authors of [10] is also consistent with the assertion that most of the produced monopoles were bound into the simplest magnetic atom, monopolium, already in a very early Universe. However, whereas a mass of monopoles $(5–10) \times 10^{12}$ GeV/$c^2$ was used in [10], most of the authors take $\sim 10^{16}$ GeV/$c^2$ for the masses of GUT monopoles (for a review, see [42]).

To the credit of our colleagues [44, 45], they noted quite long ago that at early epochs heavy monopoles were paired to form heavy monopolium, although the classic work on monopolium was performed 35 years ago [32]. It is pointed out in this paper that the lifetime of such a bound system increases cubically from the initial diameter for $SU(5)$ GUT monopoles. If the monopolium diameter is $10^{-13}$ cm, then its lifetime is only 43 days. If, however, the monopolium diameter is $10^{-9}$ cm, then its lifetime will be $10^{11}$ years. More reliable estimates were made in [44]. Here the paired primordial heavy monopoles could survive and, in the view of the authors, can be the sources of ultrahigh-energy cosmic rays above the Greizen–Zatsepin–Kuzmin (GZK) limit ($5 \times 10^{10}$ GeV) after annihilation. The size of such a bound system depends strongly on the distribution of the initial momentum $p$. An estimate of this possible size is given in [44]:

$$l_c[\text{cm}] = 5 \times 10^{-7} \left( \frac{p}{mc} \right)^4 \left( \frac{m}{10^{14} \text{ GeV}} \right) \left( \frac{10^4 \text{ GeV}}{H} \right)^2, \quad (16)$$

which lies in the range from $7 \times 10^{-7}$ to $6 \times 10^{-3}$ cm for $\Omega_x/0.3 < 1$. Here $\Omega_x$ is the relative density of the Universe and $H$ is the Hubble constant. This estimate coincides with the estimate from [32] and, probably, it may not be doubted that a bound system of heavy monopoles will survive to our days. Unfortunately,

## 7. WHY ARE MAGNETIC CHARGES NOT OBSERVED ON ACCELERATORS?

Much effort was expended [46] to detect magnetic charges with $m \approx 2.4$ GeV/$c^2$ on accelerators. However, here, probably, there was an "embarrassment", because these magnetic charges are close in mass to the vector resonance of the J/ψ particle whose mass is 3.097 GeV/$c^2$ [47]. During the J/ψ particle production its cross section is ~$10^{-31}$ cm$^2$, which is larger than the monopole production cross section, ~$10^{-32}$ cm$^2$ [40]. This "blocks" the production of the latter (of course, the interaction Lagrangian should be written out and analyzed for a rigorous assertion). Furthermore, in view of the enormous force of attraction between magnetic charges of opposite signs (stronger than that between $e^+$ and $e^-$ by a factor of 4692), they are immediately bound to form a magnetic atom—monopolium. Consequently, there have never been chances to detect Dirac magnetic monopoles with a mass of 2.4 GeV under terrestrial conditions in experiments on accelerators. They should be searched for in the cosmos, where there are no hampering effects, or in an atomic state (gamma-ray transitions during recombination), or by "catching" them in a single state on a circumterrestrial orbit. In other words, a cosmic experiment for the detection of magnetic charges is needed. In the magnetospheres of gamma-ray pulsars they, along with electron–positron pairs, can be produced and immediately bound into pairs, i.e., form monopolium. However, the stationarity equations should also be solved here to make better estimates for the concentration of magnetic charges. Note that such a problem has already been solved for $e^+e^-$ pairs by the authors of [48]. They solved the kinetic equation for the production of electron–positron pairs in pulsars with huge magnetic fields:

$$\frac{\partial F}{\partial t} + \text{div}(Fv) = Q, \qquad (17)$$

where $F$ is the distribution function of particles, $v$ is their velocity, and $Q$ is a special operator that takes into account the generation of particles by photons. The solution of this equation should be repeated for magnetic charges. Schwinger magnetic atoms with a mass of $2 \times 9.6$ GeV/$c^2$ could also be formed in the superstrong magnetic fields of pulsars, though with a lower probability, and these atoms together with Dirac magnetic atoms can be the components of dark matter, as has already been noted. From general physical considerations, there is a high probability in the formation of monopoles with small quantum numbers $k$, but whether these general physical considerations are applicable to the quantum magnetodynamics of the cosmos is still an open question.

## 8. BASIC ASSERTIONS ABOUT THE DETECTION OF MAGNETIC MONOPOLES

It follows from this paper that monopoles of various masses can exist in the Universe: heavy (relic) monopoles ($10^{15}$–$10^{16}$ GeV/$c^2$), monopoles of intermediate masses (2.4, 9.6 GeV/$c^2$), and, probably, light monopoles of leptonic masses. Of course, much effort should be expended on their detection. In 2015 researchers from the Nuclear Research Institute (Moscow) and the Joint Nuclear Research Institute (Dubna) as well as a number of Russian scientific organizations involved in the Baikal Collaboration deployed and put into operation a unique experimental cluster—the deep-water Dubna neutrino telescope on Lake Baikal, in which the detection of heavy magnetic monopoles is also envisaged [23, 24]. Now (2018) they have already three clusters under the common name "Baikal GBD—Gigaton Volume Detector". We made testable spectroscopic predictions with regard to Dirac monopoles of intermediate masses. The two-photon annihilation of para-monopolium with an energy of about 2.4 GeV in the magnetospheres of young neutron stars (gamma-ray pulsars) as a cooperative effect can be responsible for the gamma-ray excess (1–3 GeV) at the center of our Galaxy observed by the Fermi observatory. At an annihilation energy $E \approx 2.4$ GeV the energy of the ortho–para transition in Dirac monopolium, by analogy with positronium ($e^+e^-$), can be $E_{\text{ortho–para}} \approx 282$ keV, while the energy of the $L_\alpha$ transition is about 1.8 GeV. For Schwinger monopolium with a mass of $2 \times 9.6$ GeV/$c^2$ the energy of the ortho–para transition is about 1.13 MeV, while the energy of the $L_\alpha$ transition is about 7.2 GeV.

Vento's works on magnetic monopoles [28, 30, 31] are related to the possible detection of the latter at the Large Hadron Collider, but, as has already been noted, the J/ψ particle production hampers this process. Furthermore, an interesting and natural prediction in [28] is the presence of recombination radiation at the time of monopolium formation, but how to detect it at huge redshifts is a separate astrophysical problem. The question regarding light monopoles of leptonic masses is open, although the magnetic analog of the Standard Model has already been created in [20].

The interest in artificial magnetic monopoles detected in laboratories at ultralow temperatures [1–3, 13] continues to increase. Here, we are dealing with the case where the current of magnetic monopoles (zero-dimension topological defects) is observed as a result of magnetic frustration in spin ice. Actually, this is a "prosaic" situation where a macroscopic quantum phenomenon, a magnetic field source, is observed [49]. Magnetic monopoles in spin ice have already been discussed in earlier papers [50, 51], but they were called quasi-particles at that time.

Regarding Schwinger's magnetic world, it is probably necessary to accept his view [11]. The magnetic world must necessarily be realized in a very early Universe, because the magnetic interaction in this world is stronger than the electric one by a factor of 18769 and, of course, the footprint of such an asymmetry must necessarily manifest itself. No matter how we cut a magnet, its parts will always have different poles and it is impossible to get to isolated magnetic charges (monopoles), because the magnetic interaction in the present-day world is stronger than the electric one by a factor of 4692.25. At present, a magnetic atom, monopolium with a mass of $2 \times 2.4$ GeV/$c^2$, remains for spectroscopic observations of the footprints of the magnetic world, although magnetic atoms with a mass of $2 \times 9.6$ GeV/$c^2$ could also be formed in pulsars, but with a lower probability. Furthermore, there is also room for isolated magnetic charges in cosmic conditions. They can be blown out from young neutron stars, as electrons and positrons. The stationarity equations for the production, annihilation, and destruction of monopolium by gamma-ray photons should be solved here, but our knowledge for this is so far very scanty.

## 9. CONCLUSIONS

It may be pertinent to recall our view of the early Universe in order to somehow correlate the physical processes associated with magnetic monopoles. The following cosmological scenario could be realized at early epochs of evolution of the Universe: having tunneled by chance, the Universe passed from the oscillating regime to the Friedmann regime [52], probably, through a quasi-inflationary phase. Subsequently, there were leptogenesis, baryogenesis, and nucleosynthesis during its sharp cooling as it expanded. As has already been noted, Schwinger's magnetic world was formed at the epoch of leptogenesis. In the early Universe a high symmetry was lost and, of course, these processes were accompanied by phase transitions. When the symmetry was lost, light pseudo-Goldstone bosons were formed, filling the entire volume (for us these are dark matter particles). In the "dark" medium phase transitions occur as the temperature drops sharply, forming preferential scales. Baryons followed the block-phase structure prepared by phase transitions in the dark medium, forming the baryonic large-scale structure of our Universe that we observe. Note that in [53] we exploited a composite (preon) model of elementary particles, in which the presence of three generations of elementary particles is natural and which explains well some of the unsolved cosmological problems, in particular, the fractality.

Note also another important point in cosmology that is absent in our paper [54]. During its birth the Universe could probably have a larger number of dimensions and the compactification of extra dimensions must also have been of necessity. The presence of huge magnetic forces could affect the physical processes in the early Universe, but this "incantation", which is repeated in our paper in view of its exceptional importance, requires a careful study. Here we are talking about the influence of such magnetic forces on the generation of baryon asymmetry of the Universe. In Schwinger's world (early Universe) magnetic charges were immediately bound into magnetic atoms (monopolium), whence it follows that, formally, div**B** = 0 in the present-day world.

The status of experimental and theoretical research on magnetic monopoles as of 2006 was given in [55]. An older, but good review was published back in 1978 [56]. Studies of the problem of magnetic monopoles are continued with unprecedented persistence [57, 58]. Supersymmetry breaking by magnetic monopoles can be found in [59]; Kaluza–Klein monopoles and their zero modes are discussed in [60]. And there is no doubt that the initiation of studies of magnetic monopoles in the cosmos is not far off. From our viewpoint, the gamma-ray excess at the Galactic center observed by the Fermi telescope [34] can be the first footprint of the magnetic world. However, an independent spectroscopic check, for example, the detection of atomic transitions (e.g., the $L_\alpha$ line) in monopolium before its annihilation, is required here. The author of [61] proposed a different definition of the magnetic charge mass coming from the Born–Infeld electromagnetic theory [62]. As has been mentioned, the relation between the masses $m_g$ and $m_e$ in the case where the classical radii $r_g$ and $r_e$ are equal is $m_g = m_e(g^2/e^2)$ and then $m_g = 2.4$ GeV/$c^2$. However, nature could choose a different definition of the masses. As was shown by Caruso [61], the relation between the charges in the Born–Infeld electromagnetic theory is different, $m_g = m_e(g^2/e^2)^{3/4}$, and then $m_g = 0.29$ GeV/$c^2$. This point only strengthens our desire to detect magnetic charges. For completeness, note that the author of [63] thinks that all dark matter particles are magnetic dipoles consisting of two Dirac monopoles. Our estimates give a contribution of relic monopoles to the dark matter density at least at a level of 18%.

More details on the composition of dark matter can be found in our review [54]. For monopoles with energies in the GeV range a good estimate for the density cannot yet been made (the uncertainty with pulsars). Our upper limit for their density is $n_g \leq 10^{-6}$ cm$^{-3}$, for relic monopoles with energies in the TeV range it is $n_g \sim 10^{-19}$ cm$^{-3}$ [10]. For survived relic monopoles in energies $10^{15}$–$10^{16}$ GeV the density cannot differ greatly from the estimates made by Zel'dovich and Khlopov [10]—it can be only smaller.

In addition to the Baikal GVD experiment, in which it is possible to detect relic magnetic charges of very high energies ($10^{15}$–$10^{16}$ GeV), the Gamma-400 experiment that can detect individual spectral lines in magnetic atoms during their recombination is being prepared in Russia [64]. This is a complex, but very important task. In contrast, we propose to carry out a cosmic experiment to detect magnetic charges onboard the International Space Station. Quite long ago, while investigating the motion of magnetic charges, the authors of [64] defined even the unit of magnetic charge: $5\mu_B$ Å$^{-1}$, i.e., a new physical constant whose value was predicted in [17]. Probably, it makes sense to call the unit of magnetic charge "ehrenhaft" to perpetuate the name of the Austrian physicist Ehrenhaft, who apparently detected magnetic charges at the beginning of the past century (the publication on this was later, in 1942 [5]). The proposal for perpetuation was made by our "disgraced" physicist Sizov. Furthermore, the authors of [65] point out that in spin ice of dysprosium titanate (Dy$_2$Ti$_2$O$_7$) they observed a deviation from Ohm's law during the motion of magnetic charges. Note that Schwinger in his review "A Magnetic Model of Matter" [11] has already introduced the unit of magnetic charge for dually charged particles whose value is far from that announced in [65]. Furthermore, electromagnetic duality [66] in the Yang–Mills supersymmetric theory should be noted, in which a monopole condensate and even some spectrum of dions is possible. Hence the ideas of interpreting the experiments on artificial magnetic monopoles in spin ice can arise [1–3, 13].

Although our paper is devoted to searching for magnetic charges in the cosmos, we cannot but mention the present-day experiments aimed at searching for a new physics associated with the muon anomalous magnetic moment $a_\mu$, the $(g-2)$ experiment. Amaldi [29] established a lower limit for the magnetic monopole mass from this experiment 55 years ago (see Section 4). And the $(g-2)$ experiments are now continued in several laboratories worldwide: in USA (Fermilab) and Japan (KEK, J-PARC) [67]. According to Dirac, $g_\mu = 2$. It follows from the experiment that $g_\mu = 2(1 + a_\mu)$. The present-day value for $a_\mu$ is given in the review [68]:

$$a_\mu^{\text{exp}} = (11659209.1 \pm 5.4 \pm 3.3) \times 10^{-10}. \quad (18)$$

The muon anomaly has no good explanation and so far this is a hot point in physics, as are the latest experiments with spin ice [69].

There is no doubt that the detected dynamics of artificial magnetic monopoles will find a wide technological application in future (it has a conceptual significance already now). Thus, the monopole saga initiated by Dirac 87 years ago [26] unfolds with new vigor. By investigating magnetic charges and atoms consisting of them in the cosmos, we step by step enter the far and unexplored magnetic world.

## ACKNOWLEDGMENTS

I thank I.A. Ryzhkin from the Institute of Solid State Physics of the Russian Academy of Sciences for the debates on magnetic monopoles in spin ice and the referee of this review for useful remarks.

*Translated by V. Astakhov*

SPELL: 1. deconfinement, 2. skyrmions, 3. magnetricity, 4. circumterrestrial, 5. compactification